# Wide Field 3D Color Topography for Steep-slope Structures

Yuchen Lin[1#], Mengrui Wang[1#], Fan Li[2], Jingxiang Zhang[1], Manming Shu[1], Jiajing Yan[1], Chang Liu[1], Tianyu Zhao[1], Yansheng Liang[1], Ming Lei[1*]

1. MOE Key Laboratory for Nonequilibrium Synthesis and Modulation of Condensed Matter, School of Physics, Xi'an Jiaotong University, Xi'an 710049, China
2. School of Future Technology, Xi'an Jiaotong University, Xi'an 710049, China
*# These authors contributed equally to this work.*
* E-mail: ming.lei@mail.xjtu.edu.cn

Keywords: wide-field 3D imaging; ring light illumination; color topography measurement; steep-slope structures; Shape from Focus

**Abstract**

Three-dimensional (3D) topography measurement is essential for industrial inspection and scientific research. However, existing techniques are limited by an inherent trade-off between resolution, speed, color imaging capability and the ability to measure large-angle surfaces. To overcome these limitations, we propose a novel method that integrates ring light illumination based wide-field 3D imaging with an iterative shape-from-focus (SFF) algorithm. The system employs coaxial epi-illumination combined with ring light illumination to simultaneously acquire information from flat regions and steep slopes. The proposed iterative SFF algorithm achieves high-resolution 3D topography measurement. Full-color 3D reconstructions of electronic chips and PCBs demonstrate that the proposed method enables fast scanning and sub-micrometer resolution and significantly extending the measurable inclination range for large-angle structures. This method fills a critical gap in existing technology for fast, natural-color, 3D imaging and measurement of samples with steep topographies.

## 1. Introduction

Three-dimensional (3D) topography measurement has evolved from contact-based methods, such as coordinate measuring machines (CMMs), to non-contact optical techniques. Although CMMs offer high precision, their physical contact limits throughput and risks surface damage, making them unsuitable for high-efficiency inspection [1, 2]. Optical methods provide non-destructive, fast 3D topography measurement solutions and have been widely applied in industrial inspection, micro- nanofabrication, and medical fields [3-5]. Currently popular structured light projection techniques can achieve video- rate 3D acquisition over a large field of view [6-8], but their resolution is limited by the camera and projector, the resolution is limited to the tens of micrometer level. White light interferometry achieves nanometer axial resolution ideal for smooth surfaces [9-11], but is fundamentally limited by the requirement that height variations within a single pixel cannot exceed $\lambda/4$, making it unsuitable for rough or steep slope surfaces. Confocal microscopy offers sub-micrometer resolution and optical sectioning [12-15], but its scanning- dependent imaging approach results in slow speed.

Among these techniques, the acquisition of color information is often ignored, yet it plays an important role in assisting defect detection and obtaining material information. Based on wide-field microscopy, we previously developed the optical sectioning structured illumination microscopy (OS-SIM), achieving fast, micrometer-resolution 3D imaging with natural color [16-18]. However, due to the requirement of epi-illumination, OS-SIM struggles with large-angle geometries because light from steep slopes cannot be efficiently captured by the objective. Although multi-angle imaging can partially mitigate this issue [18], it incurs significant data overhead and reduces speed.

The above-mentioned microscopic topography measurement methods, including white light interferometry, confocal microscopy, and OS-SIM, all rely on epi-illumination from the objective lens to acquire information. This limits the collection of reflected light to angles constrained by the numerical aperture (NA) of the objective. When the surface slope exceeds half of this angle, imaging or measurement becomes impossible. Nevertheless, if the illumination can be delivered from a direction with a larger angle, the reflected light from steeply sloped surfaces becomes collectable. Ring light offers a potential solution for large-angle illumination. Therefore, we propose a wide-field 3D imaging technique using ring light illumination, with 3D topography reconstructed via the Shape from Focus (SFF) method. The ring light illuminates steep slopes from near-horizontal directions; scattered light from surface microstructures, defects, or roughness is captured by the objective, enabling measurement of regions that are inaccessible under conventional epi-illumination. The SFF method directly reconstructs 3D topography from wide-field imaging through focus searching [19,20]. To improve the resolution of height map under the SFF method, we developed an iterative SFF algorithm that breaks the trade-off between resolution and denoising. Our approach achieves micrometer-level resolution, fast scanning, and full-color imaging. Critically, by synergistically combining ring light illumination with SFF-based wide-field 3D imaging, we greatly extended the measurable slope range of surfaces—filling a critical gap in the existing technology spectrum.

## 2. Materials and Methods

### 2.1 Imaging system

To address the limitations of existing techniques in measuring steep-slope surfaces and acquiring color information, we developed a novel measurement system that integrates a ring light with a wide-field 3D imaging architecture. As illustrated in **Figure. 1**, a collimated white LED light source enters the microscope housing and passes through a 50/50 beam splitter. The downward-propagating beam (shown in blue) provides coaxial epi-illumination to the sample. Light reflected from the sample carries its 3D information, re-enters the objective, and is directed via mirrors to a color camera (PCO edge 5.5, Excelitas Technologies Corp., USA). A precision translation stage (PDV, with 50 cm travel in X/Y, 30 cm travel in Z; Z-axis positioning accuracy: 1 μm, China) moves the sample along the optical axis to acquire a stack of optical sections at different heights, which are subsequently fused to reconstruct a full-color 3D topography. For large-field samples, the system performs stitching by scanning multiple tiles using the XY stage capability. However, this conventional epi-illumination scheme suffers from a fundamental limitation: the maximum measurable slope is constrained by the numerical aperture (NA) of the objective, as light reflected from steep slopes fails to return into the objective. To overcome this, we supplemented the system with a customized ring light source positioned on the sample stage, consisting of 48 individually controllable white LEDs arranged in a circular pattern. When illuminating a large-angle region from a near-horizontal direction, a portion of the incident light (shown in green) undergoes scattering by surface microstructures, defects, or roughness. This diffusely scattered component partially enters the objective, enabling the system to capture information from steep slopes that would otherwise be undetectable under specular-dominant reflection. Thus, the two illumination sources play complementary roles: coaxial epi-illumination effectively captures flat regions, while the ring light provides access to large-angle regions. Their synergy yields more complete surface data than either source alone, while preserving the system's full-color imaging capability. Beyond angular coverage, our method offers advantages in imaging speed and data efficiency. Owing to its wide-field configuration, which requires only one image per axial layer, the proposed method achieves a twofold faster acquisition speed compared to OS-SIM (which requires two images per layer), while reducing data volume by half. This improvement is particularly beneficial for high-throughput industrial inspection scenarios where both speed and storage efficiency are critical.

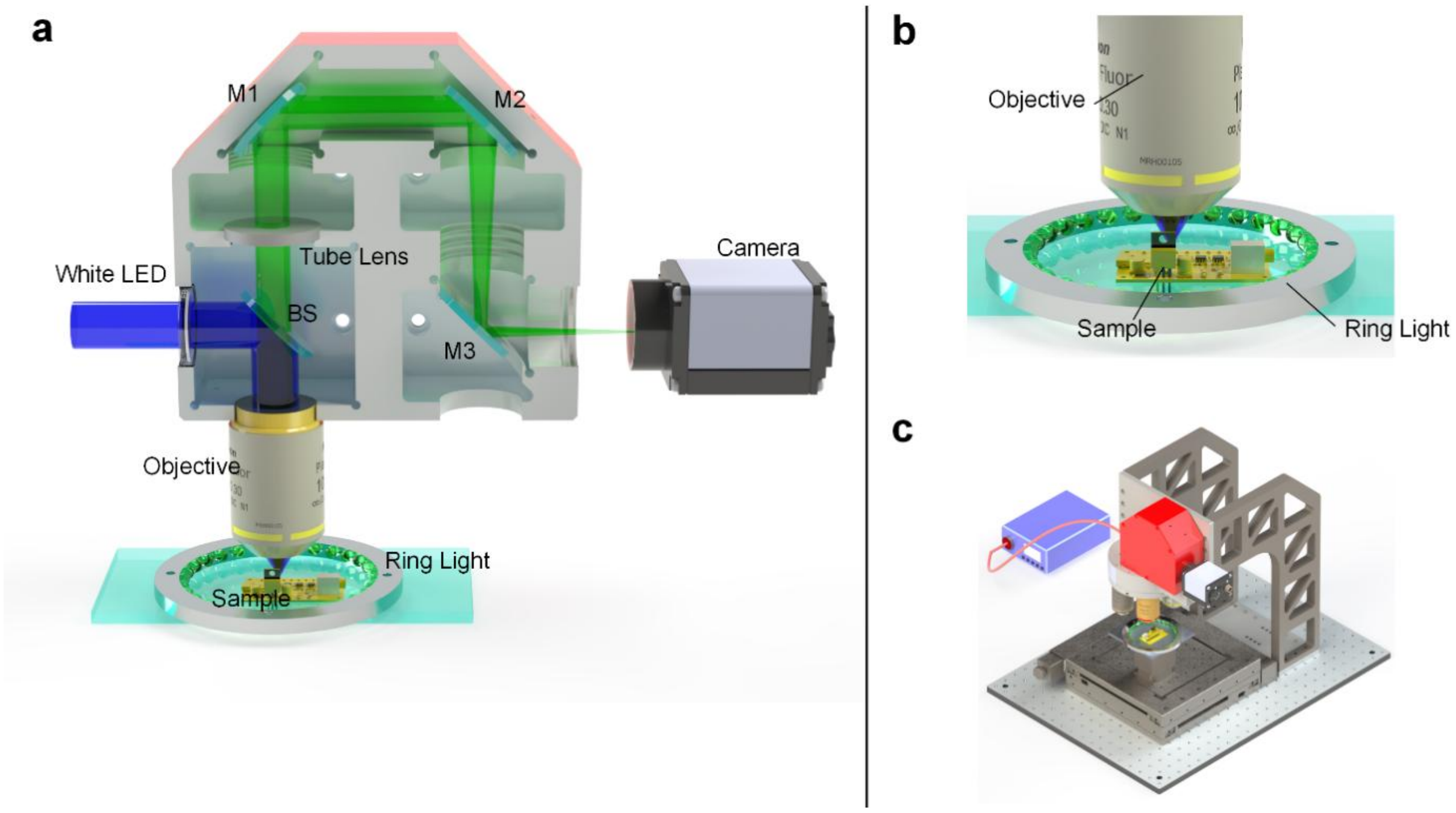

**Figure 1.** Optical path diagram and overall system diagram a. Schematic diagram of the optical path of the imaging system, where M represents the mirror, and BS represents the 50/50 beam splitter. b. Schematic diagram of the ring light c. Schematic diagram of the overall system structure.

### 2.2 Iterative shape from focus algorithm

The SFF method measures the focus level by calculating the local sharpness of WF images at different depths, and then recovers the 3D surface topography by determining the positions with maximum focus measure throughout the depth range. However, traditional SFF algorithms suffer from a trade-off between resolution and denoising: to avoid noise in the height map, the SFF operator must operate within a certain neighborhood size to establish relationships between neighboring pixels. A larger neighborhood size improves denoising but degrades the resolution of the height map, whereas a smaller neighborhood size leads to more noise. A common practice is to select a moderate neighborhood size [21], but the optimal size varies depending on factors such as sample pattern, topography, and illumination.

To break the trade-off between resolution and denoising, we propose an iterative SFF algorithm (**Figure 2a,b**), which achieves maximum resolution while maintaining denoising performance. We selected the energy of gradient (EOG) operator to calculate the focus level (FL). The number of iterations is set as $i_{max}$ (typically 5 in practice, determined by ensuring that the initially obtained height map is noise-free). At the beginning of the iteration, the FL is filtered using a Gaussian filter with the largest standard deviation $\sigma_f$. The height map $H$ is then obtained by locating the depth position where the filtered FL ($FL_{filt}$) reaches its maximum along the z-direction. Positions where the maximum $FL_{filt}$ value ($FL_{max}$) exceeds $k$ (typically 2 in practice to be regarded as significantly prominent) times the average FL are considered valid regions. The invalid regions are then padded by interpolation from valid regions, and a quadratic fitting is applied near the maximum point of $FL_{filt}$ to obtain $H_{pad}$. In each subsequent iteration, the $\sigma$ of the Gaussian filter is reduced by half, and an additional constraint is imposed when determining valid regions: the difference between $H$ and $H_{pad}$ must be less than a certain limit (gradually reduced in each iteration) to prevent the occurrence of erroneous values. After completing all iterations and obtaining the height map, the color information from the original image at the corresponding positions is mapped onto the height map to produce a 3D color image $I_{3D}$.

To validate the improvement of the proposed iterative SFF algorithm over the traditional SFF algorithm, we performed SFF-based topography measurement using a chip surface texture as the sample (**Figure 2c**). When the traditional SFF algorithm used a small neighborhood ($\sigma_f = 2$) to compute the focus level, the resulting height map contained numerous erroneous heights, i.e., noise. The neighborhood needed to be as large as $\sigma_f = 10$ tocompletely avoid the noise, but the resulting height map became overly smooth, resulting in low resolution of the surface texture. In contrast, the iterative SFF algorithm removed all noise while maintaining high resolution. As can be seen from the height profiles (**Figure 2d**), the iterative SFF algorithm eliminated all noise present in the traditional SFF ($\sigma_f = 2$) result and agreed well with it in regions free of noise, whereas the height profile of the traditional SFF ($\sigma_f = 10$) result was too smooth to reflect the texture structure.

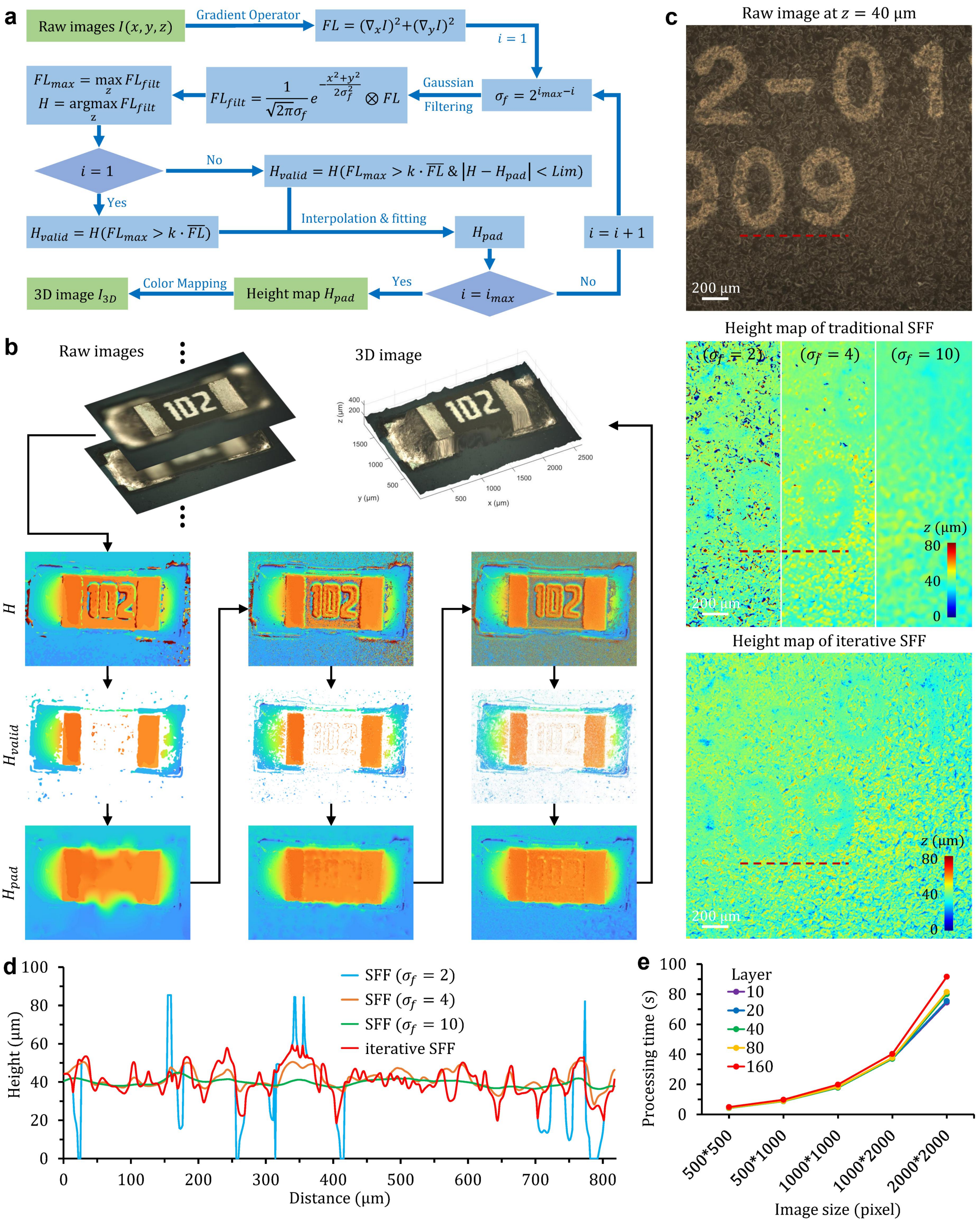

**Figure 2.** Principle and performance of the iterative SFF algorithm. a. Flowchart of the algorithm. b. Height map processing workflow illustrated using a capacitor on a circuit board as an example. c. The improvement of topography measurement through iterative SFF algorithm over traditional SFF algorithm using chip surface texture as sample. d, Height profiles along the dashed lines in c. e, Processing time under varying image sizes and numbers of layers, implemented in MATLAB on an AMD Ryzen 5 3600X CPU.

The reconstruction time of the iterative SFF algorithm mainly consists of the focus level calculation for the 3D stack and the interpolation padding of the 2D height map. The time complexity of the former is $O(i_{max}XYZ)$, while that of the latter is $O(i_{max}XY\log(XY))$, where $X$, $Y$, and $Z$ denote the image height, image width, and the number of layers, respectively. In practical processing, height-map padding implemented via natural-neighbor interpolation contributes more significantly to the total computation time. We measured the processing time under varying image sizes and layer numbers using MATLAB on an AMD Ryzen 5 3600X CPU. The results show that the processing time increases approximately linearly with image size, whereas it is not sensitive to the number of layers within a certain range (common values in practical applications). For the commonly adopted image size of 2000*2000 pixels with 160 layers, the typical reconstruction time is 91.66 s.

## 3.Results

To validate the effectiveness of the proposed method in terms of large-angle measurement capability, fine microstructure reconstruction, and full-color 3D topography imaging, we tested three types of typical industrial samples: an electronic chip, chip pins, and a printed circuit board (PCB). These samples were selected because they collectively represent the main challenges encountered in real-world automated optical inspection (AOI): namely, steep sidewalls, fine pitch structures, mixed materials, and the need for simultaneous geometric and colorimetric assessment. The experimental results, as detailed in the following subsections, demonstrate that our ring-light-assisted wide-field 3D imaging system successfully addresses each of these challenges.

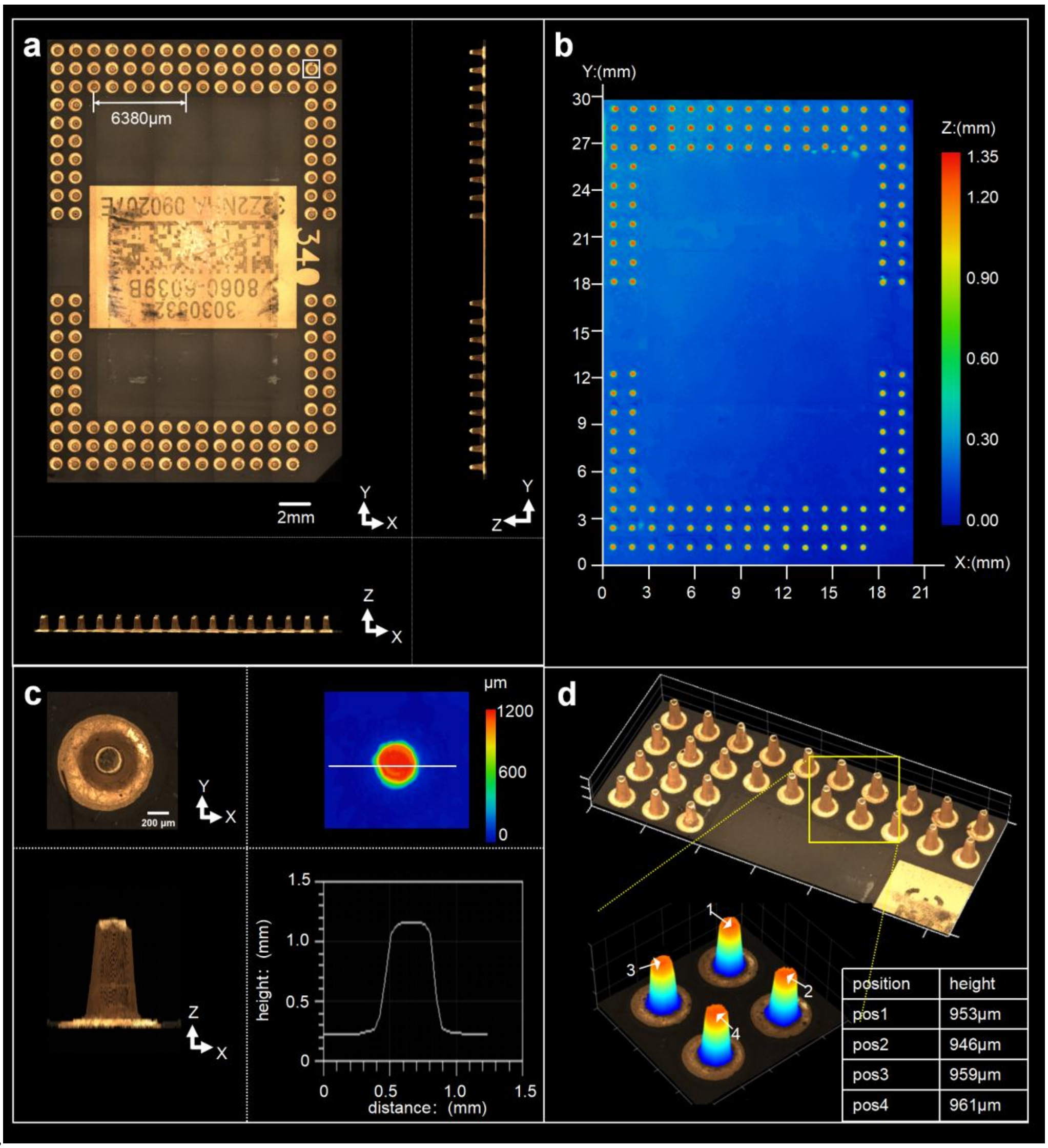

**Figure 3.** Electronic chip results. (Supplementary Video S1) a, Three orthogonal views of the electronic chip and measurement of pillar spacing. b, Height map (red-blue colormap) of the electronic chip. c, Top view of a single pin, height map of a single pin, front view of a single pin, and 3D profile measurement of an individual pillar. d, Local 3D structure of the sample with magnified view and pillar height measurement.

Accurate measurement of pillar heights on electronic chips is critical for semiconductor manufacturing process control. In particular, for heterogeneous integration and 3D chip stacking, pillar height uniformity directly impacts chip yield and packaging reliability [22]. Even minor deviations in pillar height can lead to open connections, insufficient solder wetting, or mechanical stress concentration during thermal cycling. Moreover, for advanced packaging applications such as 3D integrated packaging, bump height uniformity is a key factor in ensuring reliable interconnection [23]. Traditional stylus profilometry, while offering high vertical resolution, risks damaging these delicate microstructures due to physical contact. In contrast, our method successfully captured the full 3D topography of these pillars without contact, including their sidewall geometry (e.g., taper angle and sidewall roughness). This comprehensive dimensional data is invaluable for process optimization, failure analysis, and design validation in semiconductor manufacturing. As shown in **Figure. 3**, the electronic chip was measured using a 2× objective (NA = 0.10). A total of 8 × 5 image tiles were acquired and stitched, resulting in a total field of view of 36 mm × 23 mm and Supplementary Video S1 presents its 3D image. The system captured 160 layers at a Z-step of 10 μm. From this data stack, the 3D structure of individual pillars within the electronic chip was clearly reconstructed. Notably, precise height measurements were achieved not only for the top surfaces but also for microstructures located on the side of the chip. This capability is particularly important because many advanced packaging architectures involve multi-layer structures, where traditional top-down inspection methods fail to capture hidden features. Overall, this result strongly demonstrates the system's capability to measure fine microstructures with complex, non-planar topography.

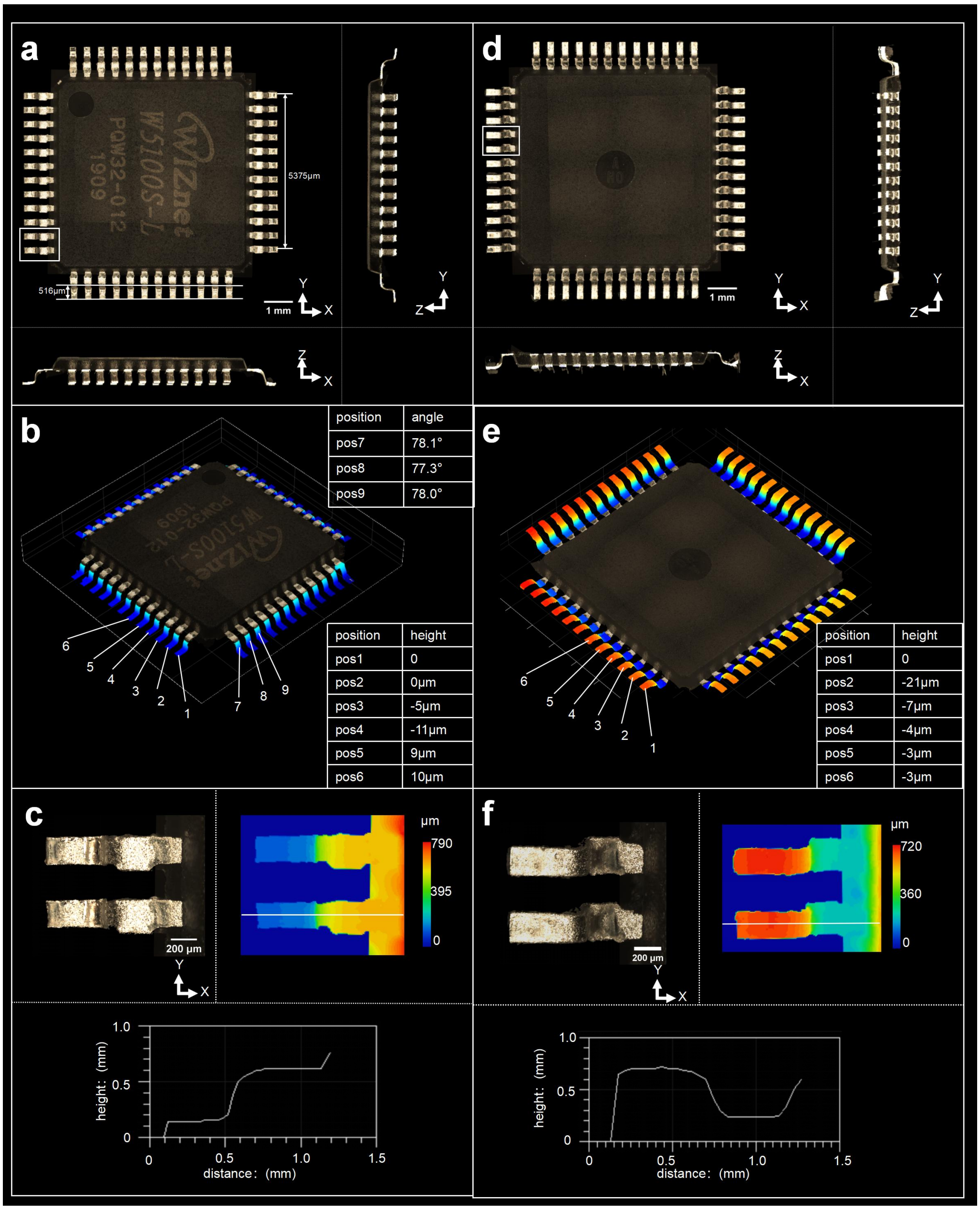

**Figure 4.** Chip pin results. (Supplementary Video S2) a, Three orthogonal views of chip pins and measurement of pin spacing. b, 3D reconstruction of chip pins with measurement of relative pin heights and angles relative to the chip's top surface. c, Top view of a local chip region, corresponding height map, and 3D profile measurement of the local chip region. d, Three orthogonal views of the backside of chip pins. e, Top view of a local region on the chip backside, corresponding height map, and 3D profile measurement of the local backside region.

Accurate measurement of IC pin geometry is essential for packaging and assembly reliability. Pin spacing, relative height, sidewall angles, and bending characteristics directly affect solder joint quality, electrical connectivity, and mechanical stability in surface-mount assembly [24]. Bent pins can cause short circuits, open circuits, or misalignment, while non-uniform pin heights lead to insufficient connections or excessive stress. Therefore, precise geometric assessment of chip pins is critical for quality control, particularly in high-reliability sectors such as automotive, aerospace, and medical electronics. To validate our method's capability for such measurements, we tested chip pins using a 4× objective (NA = 0.20). A total of 5 × 5 image tiles were acquired and stitched, resulting in a total field of view of 14 mm × 14 mm. The system acquired 250 layers at a Z-step of 5 μm, and Supplementary Video S2 presents the resulting 3D reconstruction. From this high-density z-stack, the 3D morphology of the chip pins was accurately reconstructed, and key geometric parameters—including pin length, spacing, relative height, and bending angles—were successfully measured. Remarkably, the system captured pin sidewall angles up to 78.1° (as indicated in **Figure. 4b**), a near-vertical slope unattainable under conventional epi-illumination. In conventional systems, light is delivered coaxially, so surfaces tilted beyond the collection cone angle return little signal to the detector. Consequently, bent pins with steep sidewalls often appear dark or are entirely missing from reconstructed 3D models, leading to incomplete or erroneous measurements. By employing ring-light assistance, our method overcomes this limitation. The ring light delivers oblique illumination that is preferentially scattered from steep sidewalls back toward the objective, enabling accurate 3D reconstruction without mechanical probing, multi-view stitching, or sample tilting—thus avoiding associated complexity, registration errors, and throughput penalties. This confirms that the ring-light-assisted approach significantly extends the measurable angular range, making it suitable not only for bent chip pins but also for any sample with steep features, such as micro-lenses, and machined metal parts.

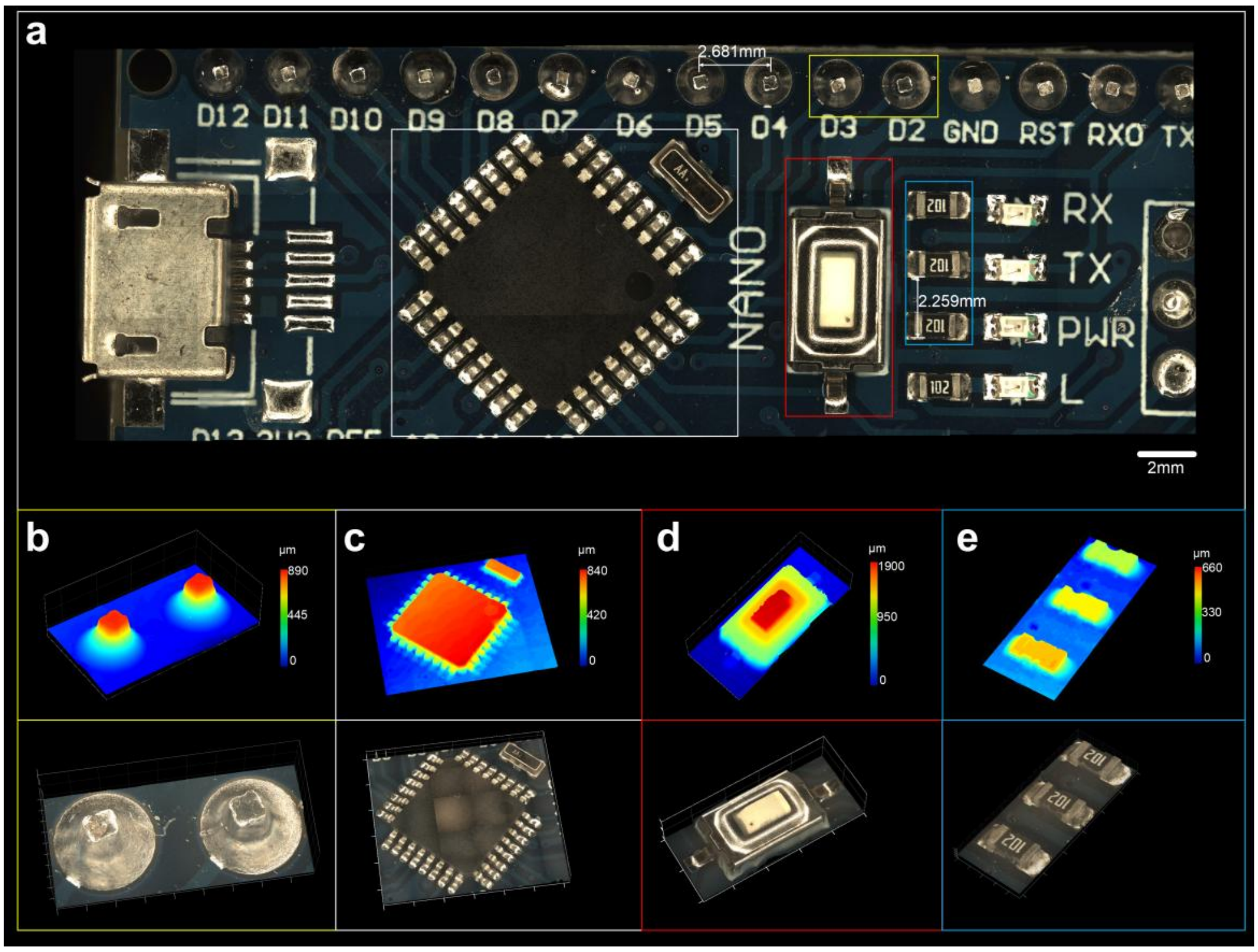

**Figure 5.** PCB results. a, Overall top view of the printed circuit board (PCB) with measurements of solder joint distances and capacitor spacing. b–e, Height maps (red-blue colormap) and corresponding 3D structures of the four local components marked in a.

PCB assembly quality directly determines the functionality and reliability of electronic products. Solder joint geometry—including wetting angle, fillet shape, and void distribution—is a key indicator of soldering process stability. Previous studies have shown that convex-shaped solder joints exhibit the best thermal shock reliability, whereas concave or insufficient fillets are prone to cracking and fatigue under thermal or mechanical stress [25, 26]. Researchers have investigated solder joint morphology primarily at the two-dimensional level [27], highlighting the need for three-dimensional, non-destructive assessment methods. Therefore, accurate 3D profiling of solder joints and electronic components is essential for process control and failure prediction in PCB manufacturing. To validate our method's capability for such measurements, we tested a PCB using a 2× objective (NA = 0.10). A total of 9 × 3 image tiles were acquired and stitched, resulting in a total field of view of 41 mm × 14 mm (**Figure 5**). The system collected 240 layers at a Z-step of 10 μm. From this data stack, the 3D profiles and surface defects of solder joints (e.g., insufficient wetting, cold joints, bridging, and voiding) were clearly revealed, while the color information of pads, solder mask, and component bodies was faithfully retained. This color retention facilitates material discrimination (e.g., copper vs. gold pads) and solder joint quality assessment based on color changes caused by oxidation or contamination. In line with previous findings, our 3D reconstruction clearly revealed the convex fillet profiles of capacitors and the characteristic wetting shapes of IC pins, enabling non-destructive, quantitative assessment of solder joint integrity. Overall, these results demonstrate the system's ability to accurately reconstruct electronic components and solder joints using 3D profiles and height maps, while simultaneously providing true color information that complements the geometric data.

**Table 1.** Imaging Parameters for Tested Samples

| | Electronic Chip | Chip Pins | Printed Circuit Board (PCB) |
|---|---|---|---|
| Magnification | 2 | 4 | 2 |
| NA | 0.10 | 0.20 | 0.10 |
| Fields | 8x5 | 5x5 | 9x3 |
| Image size | 36mmx23mm | 14mmx14mm | 41mmx14mm |
| Layers | 160 | 250 | 240 |
| Z step | 10μm | 5μm | 10μm |

## 4. Discussion and Conclusion

The proposed ring-light-assisted wide-field 3D imaging system addresses three key limitations of existing optical inspection techniques: restricted angular coverage, lack of color information, and suboptimal imaging speed. First, the integration of coaxial epi-illumination with ring light effectively overcomes the angular limitation inherent to NA-constrained systems. Coaxial illumination captures flat regions, while oblique ring-light illumination enables measurement of steep slopes—up to 78.1° in our chip pin experiment—that are otherwise undetectable. This capability extends to any sample with steep features, such as micro-lenses and machined metal parts. Second, the proposed iterative SFF algorithm resolves the trade-off between resolution and denoising inherent in traditional SFF methods. It achieves the highest resolution while maintaining noise suppression, enabling accurate reconstruction of steep surface discontinuities. Moreover, benefiting from the effective increase in NA provided by the ring light and its enhancement of surface texture details, the measurement accuracy of SFF is further improved. Third, based on WF imaging, our method offers advantages in imaging speed and data efficiency. Only one image is required per axial layer, and the measured single-layer acquisition time is 55 ms versus 80 ms for OS-SIM (which requires at least two images per layer)—a significant benefit for high-throughput industrial inspection. Furthermore, the system captures true color information simultaneously with 3D topography. This color-3D fusion capability is particularly valuable for PCB inspection, where it enables distinguishing similarly shaped components (e.g., resistors vs. capacitors), detecting pad oxidation and solder mask scratches, and reducing false positives in AOI systems. Recent research supports that color features outperform shape or texture features alone for defect detection, as many defects manifest as subtle color changes before becoming geometrically detectable [28].

Compared with existing techniques summarized in Table 2, our method uniquely combines micrometer resolution, fast scanning speed, full-color imaging, and large-angle measurement capability. White light interferometry offers excellent axial resolution but fails on steep slopes and lacks color. Confocal microscopy provides optical sectioning but is slow, NA-limited, and color-blind. Structured light projection methods are fast but have limited lateral resolution and require an additional camera for color. The OS-SIM method achieves color imaging but remains NA-limited. In contrast, our method fills this critical gap, offering a practical, non-destructive solution for semiconductor packaging inspection, PCB assembly quality control, MEMS device characterization, and precision engineering metrology.

**Table 2.** Comparison of various topography measurement techniques.

| Technique | Color | Scanning Speed | Resolution | Large-Angle Capability |
|---|---|---|---|---|
| White Light Interferometry | No | Slow | <1 nm | Limited |
| Confocal Microscopy | No | Slow | 10-100 nm | Limited |
| Structured Light projection | No | Fast | 10-50 μm | Limited |
| OS-SIM | Yes | Moderate | ~1μm | Limited |
| Proposed Method | Yes | Fast | ~1μm | Yes |

The advantage of the ring-light system over OS-SIM is most evident in angular coverage and imaging efficiency. OS-SIM performs optical sectioning via fringe modulation and decoding, which necessitates coaxial illumination and consequently restricts its collection angle to that imposed by the objective NA; multi-angle illumination only partially alleviates this at the cost of extra data and reduced imaging efficiency [17,18]. The ring-light system instead changes the illumination geometry: near-horizontal oblique illumination allows diffusely scattered light from steep sidewalls to be collected by the objective [19]. As directly compared in Figure 6 using the same 4× objective (NA = 0.20), the ring-light system captures surface inclinations up to 84.6°, whereas OS-SIM reaches only 30.6°. Specifically, the metal ring was divided into 100 segments of 3.6° each, and the maximum measurable inclination was determined by comparing the number of segments that could be captured under ring-light illumination and under OS-SIM, respectively. In addition, the proposed system requires only one image per axial layer whereas OS-SIM needs two or more phase-shifted images, giving a shorter single-layer acquisition time (55 ms versus 80 ms) and half the data volume while preserving true color. It should be noted, however, that OS-SIM retains an advantage in axial resolution, owing to its physics-based optical sectioning capability, which enables finer discrimination of axial structures. Overall, the ring-light system provides clear advantages over OS-SIM in both maximum measurable surface inclination and imaging efficiency, while the two techniques remain complementary in terms of axial resolution and angular coverage.

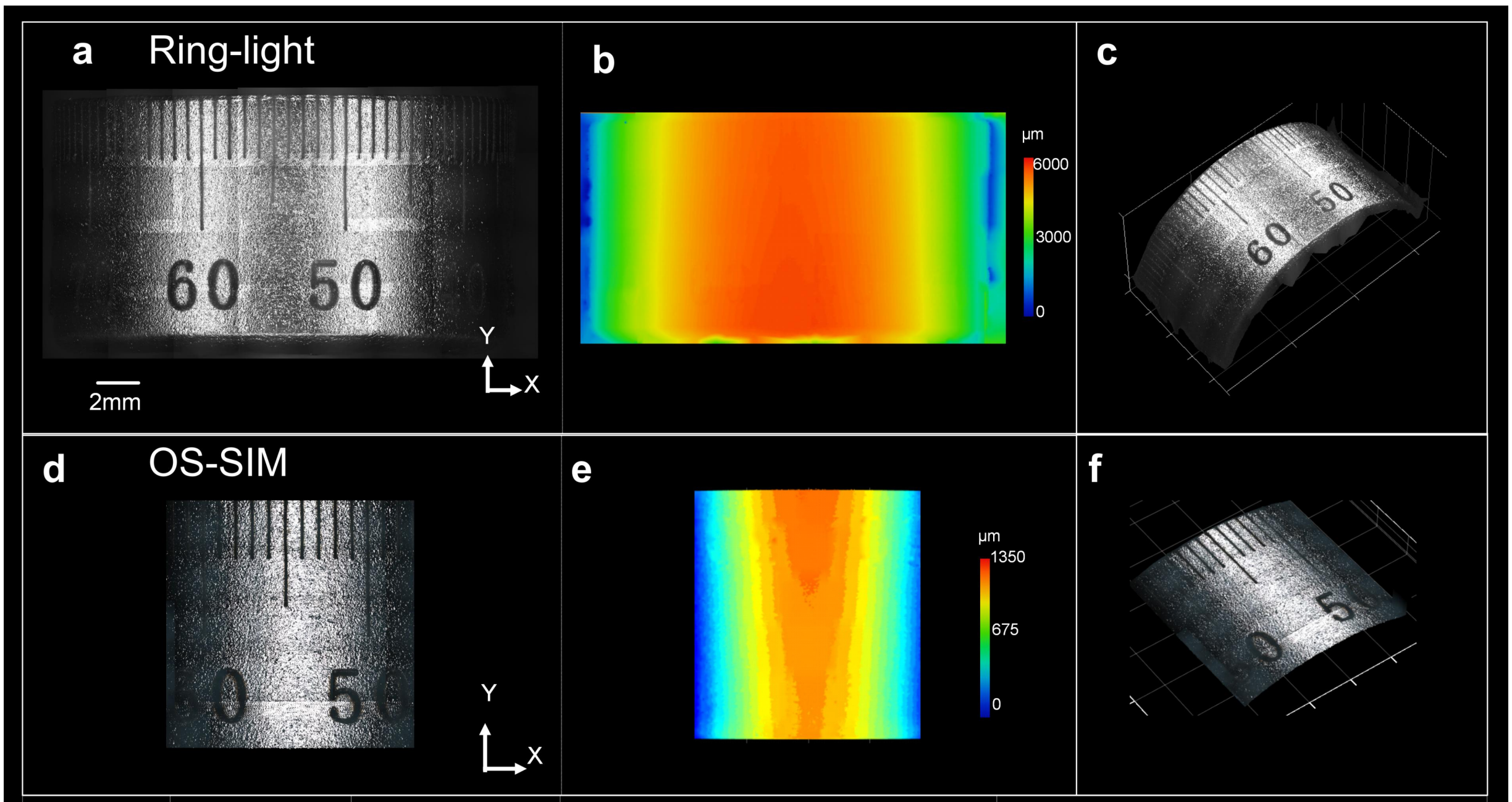

| | stage settling time | image acquisition time | total acquisition time for single layer | maximum measurable surface inclination angle |
|---|---|---|---|---|
| ring light | 30ms | 25ms | 55ms | 84.6° |
| OS-SIM | 30ms | 2×25ms | 80ms | 30.6° |

**Figure 6.** Comparison of the maximum measurable surface inclination under ring-light illumination and OS-SIM. a, Top view of the metal ring under ring-light illumination. b, Height map of the ring under ring-light illumination. c, 3D image of the ring under ring-light illumination. d, Top view of the ring under OS-SIM. e, Height map of the ring under OS-SIM. f, 3D image of the ring under OS-SIM.

Beyond angular coverage and image-number reduction, the acquisition speed is mainly limited by three factors: exposure time, stage response, and image storage. For the present system, the single-layer acquisition time is 55 ms, compared with 80 ms for OS-SIM. Brighter illumination would allow a shorter exposure time, while a larger camera buffer would reduce the image storage overhead. In addition, replacing the current translation stage with a piezoelectric ceramic actuator could shorten the stage response to approximately 5 ms. With these improvements, the single-layer acquisition time is expected to decrease to approximately 25 ms, corresponding to about 2.5 s for a 100-layer volumetric dataset. Such an improvement would further strengthen the advantage of the ring-light system in high-throughput inspection.

These speed improvements, together with the extended angular coverage, make the proposed system particularly suitable for rapid, large-area inspection of components with complex geometries. Nevertheless, we acknowledge that the current axial resolution of our method is lower than that of white light interferometry due to hardware constraints. For a large class of industrial inspection tasks that prioritize angular coverage, speed, and color information over nanometer-scale axial resolution, this trade-off is acceptable. Future work may integrate ring-light illumination with OS-SIM, where ring-light illumination ensures angular coverage for steep slopes and OS-SIM provides higher axial resolution for flat regions. This dual-modal strategy aligns with the broader trend of integrating complementary information sources for anomaly detection in manufacturing [28]. Through continued development, this technology holds promise for balancing speed, precision, and information dimensionality in both industrial inspection and scientific research, and is expected to benefit applications such as chip pins, solder joints, and micromirror arrays, where both geometric and color information are essential for effective quality control.

**Data availability**

Data underlying the results presented in this paper may be obtained from the authors upon reasonable request.

**Acknowledgements**

This work was supported by the National Natural Science Foundation of China (62135003, 62435007, 12574329); National Key Research and Development Program of China (2022YFF0712500, 2023YFF0722600); Natural Science Basic Research Program of Shaanxi Province (2024JC-YBMS494).

## Conflict of interests

Ming Lei holds a patent on the ring light illumination 3D topography measurement method (202610834440.X). All other authors declare they have no competing interests.

**Author contributions**

Yuchen Lin and Mengrui Wang contributed equally to this work. Yuchen Lin designed and built the imaging system, performed data acquisition, and drafted the manuscript. Mengrui Wang developed the iterative SFF algorithm and drafted the manuscript. Ming Lei supervised the project, conceived the research direction, and proposed the innovation. Jingxiang Zhang, Manming Shu, Jiajing Yan, Chang Liu, Tianyu Zhao, Yansheng Liang, Fan Li contributed to manuscript revision and provided experimental assistance. All authors discussed the results and approved the final manuscript.

**Supporting Information**

Supplementary information is available for this paper.